# Dynamic Imaging using any Ultrasound Localization Microscopy Dataset

Nin Ghigo[1], Gerardo Ramos-Palacios[2], Chloé Bourquin[1], Paul Xing[1], Alice Wu[1],
Nelson Cortés[3], Hugo Ladret[3,4], Lamyae Ikan[3], Christian Casanova[3], Jonathan Porée[1], Abbas Sadikot[2] and Jean Provost[1,5]

[1]Department of Engineering Physics, Polytechnique Montréal, Montréal, QC H3T 1J4, Canada
[2]Montreal Neurological Institute, McGill University, Montréal, QC H3A 2B4, Canada
[3]School of Optometry, University of Montreal, Montréal, QC H3T 1P1, Canada
[4]Institut de Neurosciences de la Timone, UMR 7289, CNRS and Aix-Marseille Université, Marseille, 13005, France
[5]Montreal Heart Institute, Montréal, QC H1T 1C8, Canada

*Abstract*— Ultrasound Localization Microscopy (ULM) relies on the injection of microbubbles (MBs) to obtain highly resolved density maps of blood circulation in vivo, with a resolution that can reach 10 μm ~ λ/10 in the rodent brain. Static mean velocity maps can be extracted but are intrinsically biased by potential significant changes in the number of MBs detected during the cardiac cycle. Dynamic ULM (DULM) is a technique developed for non-invasive pulsatility measurements in the brain of rodents, leading to temporally resolved velocity and density cine-loops. It was previously based on external triggers such as the electrocardiogram (ECG), limiting its use to datasets acquired specifically for DULM applications while also increasing the required acquisition time. This study presents a new motion matching method using tissue Doppler that eliminates the need for ECG-gating in DULM experiments. DULM can now be performed on any ULM datasets, recovering pertinent temporal information, and improving the robustness of the mean velocity estimates.

*Keywords*— Non-invasive Transcranial Dynamic Brain Imaging, Cerebral Flow, Pulsatility Index, Microbubbles, Dynamic Ultrasound Localization Microscopy

INTRODUCTION

Adequate blood flow to the brain and normal pulsatile behavior of blood vessels are essential for healthy brain function. The pulsatile behavior of a vessel is typically assessed using the Pulsatility Index (PI)[1,2] which quantifies the variations in vessel flow velocity compared to its mean velocity. PI has been widely used in patients with conditions such as diabetes, hypertension, and thyroid disorders[3]. However, the small scale of brain vasculature and the large difference in velocities range make imaging the whole brain, and obtaining robust PI measurements, challenging.

Brain angiograms, obtained through Computed Tomography or Magnetic Resonance Imaging, provide millimeter-scale anatomical maps of the brain vasculature but typically cannot capture blood flow dynamics of small vessels. Brain hemodynamics can be retrieved either in large arteries supplying the brain - with Transcranial Doppler ultrasonography[4,5] or Magnetic Resonance Imaging[6,7] – or in superficial vessels in rodents (depth < 1 mm) - with optical microscopy[8,9] or optical coherence tomography[10,11]. Accessing the blood velocity is feasible with Ultrafast Ultrasound Doppler (UFD) imaging and its use of Pulsed-Wave Doppler. UFD allows access to rapid blood flow fluctuations in large[12] and small[13] vessels.

Ultrasound Localization Microscopy (ULM)[14–16] has been developed to map microvessels of the entire brain with resolution in the 10 μm range ~ λ/10. After injection of microbubbles (MBs) in the bloodstream, each single MB is localized with a subpixel precision and tracked over time. Several minutes of acquisition are needed to reconstruct a map of the entire vascular tree and extract blood velocity markers such as the mean, minimum, and maximum blood velocities[16–20]. To avoid data overload and facilitate prolonged high frame rate imaging, a specific ultrasound sequence can be designed. Typically, ULM datasets are divided into multiple groups of frames, (usually from 200 to 600 groups[17,19,21]), each containing a constant number of frames and separated by the time needed for data storage and transfer. A conventional ULM set-up and sequence are displayed in Figure 1.ab. Since no information on when on the cardiac cycle the beginning of each group of frames is, ULM cannot capture transient and rapid phenomena beyond the extremal and average velocity measurements.

Dynamic Ultrasound Localization Microscopy (DULM) is a method that enables the non-invasive mapping of pulsatility measurements deep in the brain of rodents[22,23], the mapping of the intramyocardial vasculature of the heart[24] and, most recently, the functional activity of the brain[20]. Ultrasound acquisitions are typically gated on an external signal such as the animal electrocardiogram (ECG) and/or breathing, allowing for a coherent temporal summation to obtain a dynamic cine-loops of the vascular tree. However, external triggers have several downsides, the main one being that the acquisition must be stopped periodically to wait for the trigger. This increase in wait time leads to fewer MBs present in the bloodstream due to their limited lifespan.

Herein, a motion-matching algorithm based on tissue Doppler is presented, eliminating the need for ECG-gating and enabling dynamic cine-loops to be extracted from any ULM dataset. We show that this approach enables the generation of blood flow dynamics cine-loops along with mean velocities unbiased by the variation of the number of detected MBs during the cardiac cycle all while decreasing DULM acquisition duration and experimental complexity. We also present maps of the Pulsatility Index in the whole brain.
We first validated the motion-matching algorithm on a gated DULM dataset of a rat brain with craniotomy, where a random additive temporal delay was introduced to desynchronize the acquisition. The method was then applied in 2D in a mouse brain through the skin and skull, as well as in 3D in a cat brain with craniotomy. In both animals, cine-loops of MBs density and velocity could be visualized. More accurate mean velocities were recovered with DULM compared to with standard ULM with a deviation as high as 6% of the maximum flow velocity in highly pulsatile vessels.

## MATERIALS AND METHODS

### A. Experimental Procedure

The experimental procedures in this study were conducted following the guidelines provided in the guide for the care and use of laboratory animals by the Canadian Council for Animal Care.

Ultrasound imaging was performed using a Vantage imaging system (Verasonics, WA, USA). A 128-elements, 18-MHz linear array probe (L22-14 probe, Vermon, France) and a 32-by-32-element, 8-MHz 2-D matrix-array probe (Verasonics, WA, USA) were used for 2D and 3D imaging, respectively. Freshly activated microbubbles ($1.2 \times 10^{10}$ MBs per mL, Definity, Lantheus Medical Imaging, MA, USA) were diluted in saline and injected as a bolus depending on the weight of the animal.

Experiments were conducted in 1) a rat with craniotomy and ECG-gating in 2D, 2) three mice in 2D through skin and skull, and 3) a cat with craniotomy in 3D (Table I).

*1) 2D Rat Experiment*
A female rat was anesthetized with 2% isoflurane and underwent a craniotomy. The rat body temperature was maintained at 35°C throughout the experiment using a small animal monitoring platform (Labeo Technologies Inc., QC, Canada). This platform also recorded the ECG and gated the ultrasound system on the R-wave of the cardiac cycle. The rat head was secured on a stereotaxic frame. The ultrasound acquisition was started 30 seconds after a bolus injection of 50 μL MB solution diluted in 50 μL of saline.

*2) 2D Mouse Experiment*
Three male mice were anesthetized with 2% isoflurane, and their heads were secured on a stereotaxic frame. Throughout the acquisition, the mice temperature was maintained at 35°C with a feedback-controlled heated blanket. An 80-μL MBs solution diluted in saline (1:10) was injected just before the ultrasound acquisition.

*3) 3D Cat experiment*
A detailed description of the procedure can be found in [25]. Briefly, after anesthesia (1.5% isoflurane), a solution of 2% gallamine triethiodide was used for immobilization and a tracheotomy was performed for artificially ventilation. A 15 × 15-mm² craniotomy window was performed on the right side of the cat brain. Throughout the acquisition, body

### TABLE I
### ANIMALS PREPARATION

|  | Rat | Mice (n=3) | Cat |
|---|---|---|---|
| *Sex* | Female | Male | Female |
| *Surgery craniotomy* | (~15 mm 10 mm window) | None | (~15 mm 15 mm window) |
| *Microbubbles solution injection* | In the jugular vein 100 μL diluted at 1:1. | In the tail vein 80 μL diluted at 1:10 | In the paw 1 mL diluted at 1:1 |
| *ECG gating* | On the R-wave | None | None |
| *Ethics Committee* | Montreal Heart Institute (2019-2464, 2018-32-03). | Animal Use Protocol (AUP-4532) | University of Montreal (CDEA 19-008) |

### TABLE II
### SEQUENCE PARAMETERS

|  | 1 Rat | Mice | 1 Cat |
|---|---|---|---|
| *Probe* | 18-MHz 128 elements | 18-MHz 128 elements | 8-MHz 32 × 32 elements |
| *# Groups acquired* | 800 | 400 | 800 |
| *# of transmits / receives* | 3 per frames (-1°,0°,1°) | 11 per frames (-5/5°,-4/4°, -3/3°,-2/2°, -1/1°,0°) | 4 per frame 0° |
| *Transmit pulse* | 3 cycles at 15.625 MHz | 3 cycles at 15.625 MHz | 4 cycles at 6 MHz |
| *Frame rate (Hz)* | 1000 | 1000 | 1000 |
| *# Frames per group* | 400 | 600 | 600 |

temperature was maintained at 37°C with a feedback-controlled heated blanket. A 2-mL MBs solution diluted in saline (1:1) was injected just before the ultrasound acquisition.

### B. Bias in mean velocity estimator in ULM

In ULM, mean velocity measurements are biased due to the inability to account for variations in the number of detected microbubbles (MBs) during the cardiac cycle. Typically, the MBs positions and velocities from all groups of frames are accumulated on a grid without considering any temporal information, resulting in a static vascular map of the brain MBs density and summed velocities. The mean velocity map is then obtained by dividing the summed velocities by the density map[18–20]. This approach introduces bias in the average velocity map, especially in highly pulsatile vessels where the number of detected MBs can significantly differ between diastole and systole.

To reduce bias, the mean velocity estimator should be weighted by the number of MBs detected at each instant in a given pixel, denoted as $N_{MB}$, as typically done in Dynamic ULM:

$$\bar{v} = \sum_{t=0}^{t=NT} \left( \sum_{n=1}^{n=N_{MB}(t)} v_{MB}(n,t) \Big/ N_{MB}(t) \right) \quad 1)$$

where $\bar{v}$ represents the mean velocity, $T$ the duration of a cardiac cycle, N an integer and $v_{MB}$ the velocity of the observed MB.

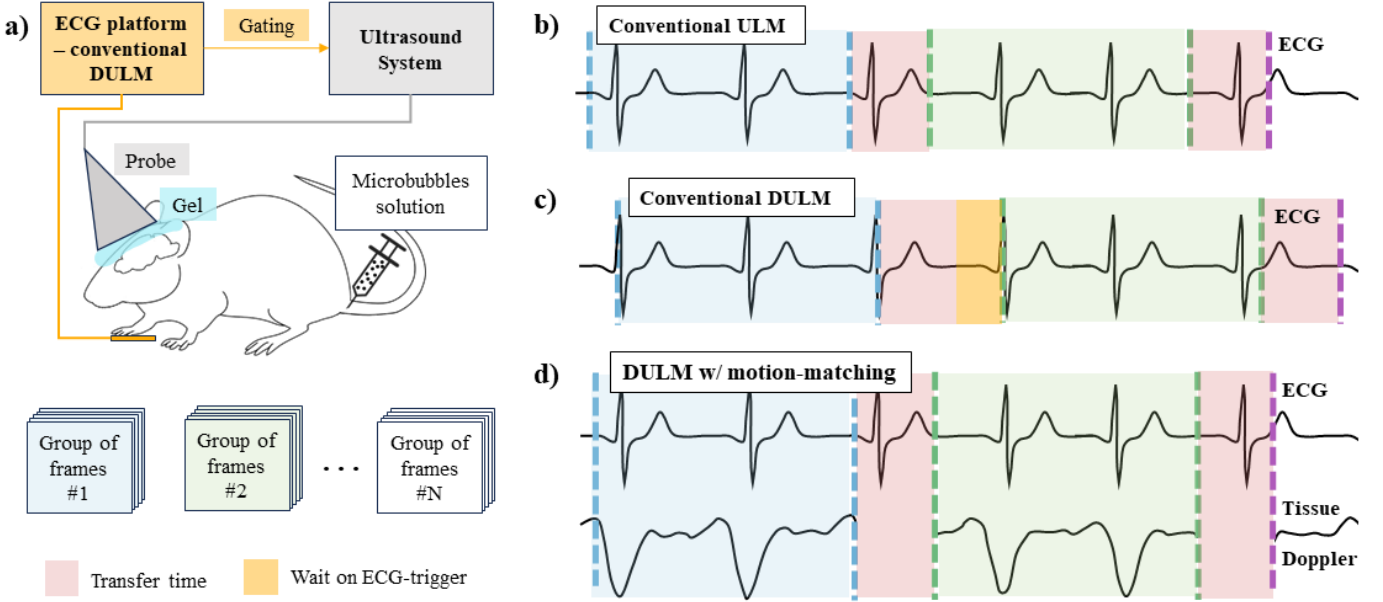

Figure 1: **Differences between conventional ULM and conventional and motion-matching DULM**. (a) Illustration of a conventional DULM experimental set-up, i.e., with the use of an ECG-trigger. Both conventional ULM and motion-matching DULM set-ups are similar, differing only in the absence of the ECG-platform. After the microbubble injection, the ultrasound sequence captures the microbubbles circulating through the bloodstream of the animal, dividing the dataset into multiple groups with a consistent number of frames. Subsequently, (b), (c), and (d) depict the temporal events of conventional ULM, conventional DULM, and DULM with time-registration by motion-matching acquisition, respectively, using representative animal ECG signals. The ECG signal is only available in conventional DULM, where an ECG-platform is used to trigger the ultrasound system. As an example, only the two first groups of frames of the acquisition are represented in blue and green, while the time needed for data transfer is highlighted in red. (b) In conventional ULM, the groups of frames start at random points in the ECG signal. (c) In conventional DULM, the groups of frames start at a specific point in the ECG fixed by the ECG-trigger. The time spent waiting for the trigger is highlighted in orange. (d) In DULM with motion-matching, the sequence is similar to conventional ULM. The groups of frames can be later synchronized using the cyclic properties of the tissue Doppler.

However, the commonly used mean velocity estimator in ULM is typically:

$$\bar{v}_{ULM} = \frac{1}{N_{tot}} \sum_{n=1}^{n=N_{tot}} v_{MB}(n) \quad \quad 2)$$

where $N_{tot} = \sum_{t=0}^{t=NT} N_{MB}(t)$, the total number of MBs detected and where all temporal information is discarded.

Equation 1 and 2 are equivalent only if the number of detected MBs is constant over time, i.e., if $N_{MB}(t)$ is a constant function.

### C. Dynamic ULM with Motion Matching Time Registration

*1) Ultrasound Sequence*

A compound plane wave sequence of either 1 (cat), 3 (rat) or 11 (mice) angles spaced at 1° intervals was used to acquire 800 (rat/cat) or 400 (mice) continuous groups of 400 (rat/cat), 600 (mice) frames at a frame rate of 1000 Hz. A 15.625-MHz, 3-cycle-long pulse was used for rodent brains and 6-MHz, 4-cycle-long pulse was used in the cat brain. Other ultrasound parameters are reported in Table II.

*2) DULM processing*

The DULM processing employed in this study followed a similar approach to the method described in[22], with the addition of aberration correction[26] for the acquisition through the mouse skull.

The acquired IQ data were beamformed using a delay-and-sum algorithm, with a f-number of 1, on an isometric grid with a pixel size of 23.5 μm i.e., λ/4 (rat/mouse) and 128 μm i.e., λ/2 (cat). A spatio-temporal clutter filter based on Singular Value Decomposition was then applied to suppress stationary signals and retain the moving microbubble (MB) signals[27].

Depending on the intensity of the tissue signal, between 10 to 25 singular values were removed, corresponding to the best qualitative trade-off between preserving slow MB signals and canceling tissue signal. Local maxima were detected on the IQ filtered images and considered as potential MBs. The correlation between a potential MB and the Point Spread Function (PSF) of the imaging system was used as a confidence metric. The PSF was obtained through simulation using the software SIMUS[28]. Each potential MB associated with a correlation value higher than 0.4 (rat/mice) or 0.3 (cat) was considered a true detection, while others were rejected as false detection. The sub-pixel position of each MB was estimated using a Gaussian fitting algorithm over a 7-pixels spatial neighborhood and a time ensemble of 7 ms. MBs were then tracked with the Hungarian tracker implemented by[29], with no gap filling and a maximum linking distance of 2 pixels between consecutive frames. Cubic smoothing spline interpolation was performed on each track, and the derivative of the spline was extracted as the MB velocities.

To correct for aberrations introduced by the mouse's skull, a complex-valued convolutional neural network (CV-CNN) as described in [26] was used. Briefly, the CV-CNN was trained on simulated radiofrequency signals corresponding to MBs trajectories using 20,000 different aberration functions. The network was then fed the in-vivo estimated MBs tracks longer than 16 frames from the last 100 groups of frames of the mouse acquisition. This process provided a pixel-wise phase aberration function, which was subsequently incorporated into the beamformer as time delays for phase aberration correction. The DULM pipeline was then reapplied to the newly beamformed IQ data.

This DULM process was repeated for each group of frames. Given the potential for motion occurring between these frame groups, stemming from normal experimental dynamics, a rigid motion compensation technique was employed. This involved computing correlations between the power Doppler of each frame group and a selected reference group. In the Fourier domain, the maximum of this correlation indicated the estimated pixel displacement between these groups. The positions of MBs associated with each group were then adjusted based on this estimated motion. By accumulating the positions and velocities of MBs over time and groups of frames and projecting them onto a highly resolve 10 μm ~ λ/10 (rat/mice), 50 μm ~ λ/4 (cat) isometric Cartesian grid, precise vascular maps of MB density and mean velocity within the brain could be obtained.

### 3) Motion Matching Time Registration

For each group of frames, MBs positions and velocities were recovered using DULM processing. When the acquisitions were synchronized with the ECG, the beginnings of each group of frames were already aligned with the cardiac cycle of the animal (Fig. 1c). This allowed for direct averaging of MBs velocities across the groups of frames, enabling the observation and quantification of dynamic phenomena.

If no ECG-gating is used, the beginning of each group of frames does not coincide with the same time in the cardiac cycle (Fig. 1b). As a result, direct averaging of MBs velocities is not feasible. Time registration is hence necessary to synchronize each group of frames with the cardiac cycle (Fig. 1d). We developed a motion-matching algorithm[30] for time registration by leveraging the cyclic tissue motion of the brain. Tissue Doppler was estimated for each group of frames within multiple locations in the imaging volume, using a 5-pixel spatial neighborhood and a time ensemble of 30 ms. Locations of interest were identified as the local maxima in the skin above the skull. For motion matching, the correlation between the tissue Doppler signals of a reference group and the other groups of frames was computed. MBs tracks were delayed by the respective maximum correlation value associated to each group of frames. As a result, each group of frames are synchronized with the brain tissue motion of the reference group, and consequently, with its cardiac cycle. MBs velocities can be averaged to retrieve the flow velocity variation through time and extract dynamic markers.

### D. Pulsatility index (PI)

#### 1) PI calculation

The PI is defined as the difference between the peak systolic velocity (PSV) and the end diastolic velocity (EDV), divided by the mean flow velocity (MFV):

$$PI = (PSV - EDV)/MFV \quad 1)$$

#### 2) PI Mapping

Cine-loops of the MBs density and velocity were obtained on a highly resolved 10 μm ~ λ/10 (mouse), 50 μm ~ λ/4 (cat) isometric grid. For each pixel, the PSV and EDV values were considered as the 95th and 5th percentiles over time, respectively. By utilizing percentiles instead of extrema values, this method reduced sensitivity to noise and outliers and provided smoother PI map. A median filter was applied to the resulting PI map within a 5-pixel neighborhood. Since the robustness of MVF, and consequently PI, is influenced by the number of MBs detected in each pixel, a density threshold of 25 MBs was applied to display only relevant PI values.

#### 3) Vessel wise velocity and PI

While obtaining a highly resolved PI map is beneficial for easy visualization and interpretation, caution should be exercised when using it for further computations, such as extracting the PI of a specific vessel of interest. This is due to the introduction of grid effects when projecting sub-pixel positions onto a grid, which decreases the accuracy of velocity mapping and subsequent computations.

In this study, we proposed to use the static highly resolved density map to segment vessels and extract all hemodynamic information using the sub-pixel MBs tracking data. Vessels of interest were manually segmented and for every frame, MBs tracked within the vessel mask were extracted. Their velocities were summed along time and the mean temporal velocity could be recovered by dividing by the number of MBs detected at each frame.

The velocity signals were then smoothed using a low-pass Butterworth filter of the 3rd order with a cutoff frequency selected based on the animal heart rate (35, 30, and 20 Hz for the mouse, rat, and cat respectively). The vessel-wise PI could then be obtained using Eq 1). To enhance robustness, the final PI was averaged over all the cardiac cycles available.

### E. Vessel segmentation and diameter estimation

#### 1) In 2D

First, the direction of the vessel was extracted as the average of the track direction at its center. A highly resolved map of the beam-to-flow direction, denoted $\theta$, can be obtained using:

$$\theta(x,z) = \frac{1}{N_{tot}(x,z)} \sum_{n=1}^{n=N_{tot}(x,z)} \sin^{-1} \frac{\Delta x(n)}{\sqrt{\Delta x(n)^2 + \Delta z(n)^2}} \quad 4)$$

where $N_{tot}(x,z)$ it the total number of MBs flowing through the super-resolved pixel of coordinates $(x,z)$ ; $\Delta x$ and $\Delta z$ are defined for any MB $n$, except the last one in its track, as $\Delta x = x(n+1) - x(n)$ and $\Delta z = z(n+1) - z(n)$ where $n+1$ indicates the subsequent MB in the track.

Vessels of interest were manually segmented, and the centers of these vessels were considered as the locations with the highest MB density within the segmented region. For each pixel at the center of the vessel, the vessel diameter was estimated on the MB density cine-loop as the full width at half maximum of the perpendicular cross-section to the associated beam-to-flow direction. A final temporally resolved diameter was then obtained for each vessel by averaging the temporally-resolved diameters found at all the centered pixels.

#### 2) In 3D

First, a Gaussian filter was applied (sigma = 100 μm, kernel of 3 x 3 x 3 voxels) to the 3D static density map. Vessel segmentation was performed on the resulting map using a Hessian filter. Radius was then computed for every centreline point in all detected vessels.

## F. Motion Matching Time Registration Validation

To validate the time registration method, the ECG-triggered rat acquisition was employed. To simulate an acquisition without ECG-gating, a random additive delay ranging from -100 and 100 ms was introduced to each group of frames. The motion matching time registration was then applied to resynchronize the groups of frames. The velocity maps extracted before the introduction of delays, and after the introduction of delays and time registration should yield comparable results. For the mouse and cat acquisitions no ECG-gating was used. The application of motion matching time registration should enable the visualization of distinct cardiac cycles and pulsatile phenomena that would be otherwise unobservable.

## RESULTS

### G. DULM with motion matching time registration allows for the extraction of dynamic phenomena on any ULM dataset

*1) Motion Matching Time Registration Validation in the rat brain*

Fig. 2.a displays a static map of the rat brain intensity coded reflecting the sum of microbubbles (MBs) in each pixel (10μm×10 μm, ~ $\lambda/10 \times \lambda/10$). The vasculature is well represented, with MBs flowing downward shown as positive (red) and MBs flowing upward shown as negative (blue). Complex vessels anatomy can be observed, with vessels flowing upward and downward intricately linked. In Fig. 2.c, the mean velocity map is displayed on a similar grid as the density map. Several vessels of interest of different sizes and flow directions were manually segmented and are highlighted by white rectangular boxes. In Fig. 2.e, the velocity detected within these vessels is displayed over time (400 ms). Two full cardiac cycles are distinctly discernible on the ECG-gating acquisition. This temporal information is completely lost in the simulated acquisition without ECG-gating but can be successfully retrieved through the application of motion-matching time-registration.

*2) Motion Matching Time Registration on the mice brain*

Fig. 2.b displays a static map of a mouse brain intensity coded reflecting the sum of MBs in each pixel (10 μm×10 μm ~ $\lambda/10 \times \lambda/10$). Complex vessels anatomy can be observed, with vessels flowing upward and downward intricately linked. In Fig. 2.d, the mean velocity map is displayed on a similar grid as the density map. Several vessels of interest of different sizes and flow directions were manually segmented and are highlighted by white rectangular boxes on the density map. In Fig. 2.f, the velocity of these vessels over time (400 ms) are plotted, and three full cardiac cycles can be visualized. Vessel-wise pulsatility index (PI) could be extracted with a large range of value, from 0.12 to 0.79. The PIs were extracted and averaged on each cardiac cycle, and are reported on Table III, along with the mean velocity and the diameter of each vessel.

*3) Motion Matching Time Registration on a cat brain in 3D*

Fig. 3.a displays a static map of the cat brain MBs density reflecting the sum of MBs in each pixel (50 μm×50 μm ~ $\lambda/4 \times \lambda/4$). The vasculature is well represented, with vessels of different diameters and tortuosity. In Fig 2.b, the mean velocity map is displayed on a similar grid as the density map. In Fig. 3.c, the temporal-velocities of segmented vessels, highlighted by colors on the density map, are plotted, and one full cardiac cycle can be visualized over the 600 ms. Vessel-wise Pulsatility Index (PI) could be extracted with a large range of values, from 0.05 to 0.09, and are reported on Table IV, along with the mean velocity and the diameter of each vessel.

### H. DULM leads to more robust mean velocity maps than ULM

Averaging MBs velocities over a set of full cardiac cycles after motion-matching time-registration can consider the varying number of MBs detected during the cardiac cycle's different phases. The difference between the average velocities obtained through DULM and the ones obtained through standard ULM, i.e., without synchronizing groups of frames, is displayed in Fig 4. for one mouse. Depending on the variation in the number of MBs detected during the cardiac cycle, and the velocities, the averaged velocities difference between DULM and ULM varies between 0 and 3% of the averaged DULM velocities (-1 and 1 mm/s). This bias was also studied at the vessel scale (Fig 4.cd). Several vessels of interest, highlighted by white rectangles in Fig 4.a, were manually segmented and the mean velocity obtained with conventional ULM and DULM were extracted. When the variation of MBs density during the cardiac cycle is important compared to the average density, (Fig 4.c), the bias ULM inherently possesses is also important. However, when the variation of MBs density during the cardiac cycle stay relatively small compared to the average density, (Fig 4.d), the mean velocity recovered with ULM and DULM is similar.

TABLE III
DULM ALLOWS FOR ESTIMATION OF MARKERS OF INTEREST IN A MOUSE BRAIN

| VESSEL | DIAMETER ± STD (μM) | MEAN VELOCITY (MM/S) | PULSATILITY INDEX ± STD |
|---|---|---|---|
| 1 | 44±5 | 42.2 | 0.14±0.08 |
| 2 | 51±3 | 40.9 | 0.22±°0.07 |
| 3 | 62±7 | 47.8 | 0.12±0.02 |
| 4 | 49±4 | 41.2 | 0.17±0.07 |
| 5 | 40±5 | 52.1 | 0.79±0.10 |

TABLE IV
DULM ALLOWS FOR ESTIMATION OF MARKERS OF INTEREST IN THE CAT BRAIN IN 3D

| VESSEL | DIAMETER (μM) | MEAN VELOCITY (MM/S) | PULSATILITY INDEX |
|---|---|---|---|
| 1 | 360 | 42.2 | 0.06 |
| 2 | 332 | 40.9 | 0.08 |
| 3 | 412 | 47.8 | 0.07 |
| 4 | 400 | 41.2 | 0.09 |
| 5 | 400 | 52.1 | 0.05 |

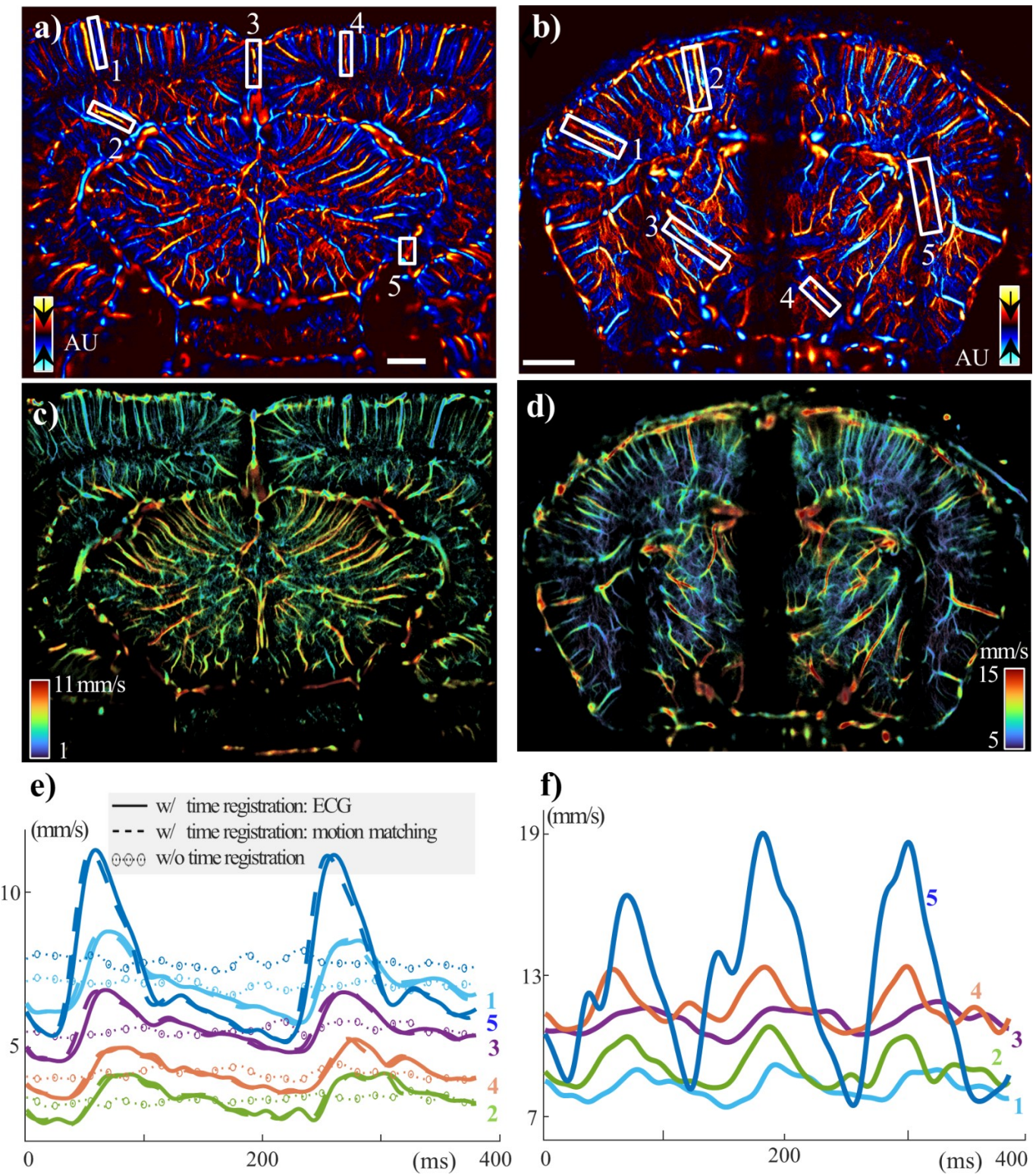

Figure 3: **Velocity measurement in the rat brain with craniotomy and in the mouse brain**. (a-b) Density based rendering of the microbubble summation for the rat and the mouse respectively (red: flowing downward; blue: flowing upward). The map resolution is 10 μm × 10 μm ~ λ/10×λ/10, the white scale bar represents 1 mm. (c-d) corresponding velocity maps for the rat and the mouse respectively. (e) Velocity variations over time (400 ms) of vessels of interest highlighted by white rectangles in (a) for the rat. The solid lines are obtained directly on the ECG-gated acquisition. The dashed lines are obtained after introduction of random delays on the ECG-gated acquisition and after performing time registration. The dotted and circled lines are obtained after introduction of random delays without time registration. (f) Velocity variations over time (400 ms) of vessels of interest highlighted by white rectangles in (b) for the mouse.

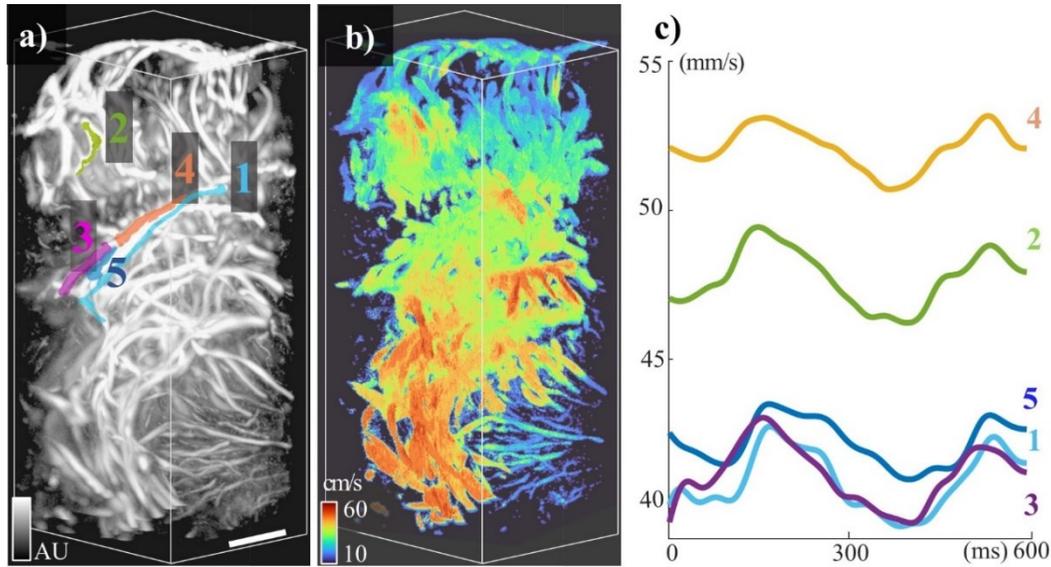

Figure 4: **Velocity measurement in the cat brain in 3D**. (a) Density based rendering of the microbubbles summation. The map resolution is 50 μm×50 μm ~ λ/4×λ/4, the white scale bar represents 1 cm. Several vessels of interest are highlighted by a color overlay. (b) corresponding mean velocity map (c) Velocity variations over time (600 ms) of vessels of interest highlighted by a color overlay in (a).

### I. Pulsatility Index Maps

Fig 5.a displays the PI map obtained for the three mice brain. A large range of PI can be observed with value ranging from 0.1 to 1.4. On the whole coronal slice of the brain, the distribution of PI follows a bimodal distribution as shown the histogram distribution displayed in Fig 5.b. Veins and arteries were separated in the manually segmented cortex using the direction of the tracks – veins flowing upward and arteries flowing downward. The PI of each pixel corresponding to arteries or veins could hence be compared in the cortex, showing two different PI distributions (Fig 5.c). A minimum of 100 MBs was used as a threshold on each pixel to reduce the noise.

### DISCUSSION

In this study, the advantages of Dynamic Ultrasound Localization Microscopy (DULM) were studied compared to conventional Ultrasound Localization Microscopy (ULM). Moreover, a motion matching method was proposed, allowing the obtention of DULM with any ULM datasets. DULM allows not only to visualize and study dynamic flow phenomena, such as the Pulsatility Index (PI), but also leads to more robust mean velocity estimates by considering the variation of detected microbubbles (MBs) during the cardiac cycle. DULM could now be performed in place of ULM, even for the study of static mean velocity maps.

Without the need of any prior information on the cardiac cycle of the animal, such as an ECG-gating, DULM was successfully performed, leading to complex density and velocity cine-loops in both the mice brain in 2D and cat brain in 3D (Fig 2 and 3). A PI map was extracted in the whole brain of three mice and low and high PI vessels could be recovered. In vessels where few MBs were detected, PI map extraction is still limited by its need for division by the mean velocity. The PI map should hence be used as a visualization tool, and vessels of further interest should be segmented for smoother PI estimation. Vessel-wise PI estimation was performed on several vessels of interests with different flow direction, diameter and at different depth, highlighting the robustness of DULM for velocity estimations. In the case of the cat, we observed relatively low PI values for the segmented vessels (Table IV). This phenomenon may be attributed to the prolonged anesthesia administered to the cat (7 days), which can induce vessel dilation, resulting in reduced diameter and blood flow variation during the cardiac cycle[31,32]. Nevertheless, even under these challenging conditions, we successfully applied DULM and recovered temporal velocity waveforms in alignment with the cat heart rate (Fig 3.c).

A Complex-Value Convolutional-Neural-Network[26] was used for aberration correction. Even with the presence of a mouse skull and skin, DULM was successfully applied, and complex phenomena studied (Fig 2,5). This step is particularly crucial for better MBs positions estimation, and hence tracking and velocity estimation.

While the tracking methodology utilized in this study is commonly employed in both ULM and DULM applications[33,19,34], it's worth noting that tracking techniques in the field of ULM have advanced significantly, yielding more robust trackers. Implementing these advanced trackers is advisable, especially when investigating dynamic phenomena[23,35,36].

DULM relies on the assumption that phenomena of interest remain constant over a few cardiac cycles (typically between 2 and 5 depending on the heart rate, from 0.3 to 1 second). This is a common assumption that holds in many cases[37–40]. However, under severe heart diseases, such as ventricular hypertrophy or myocardial infarction, this can be untrue as the QRS model tends to deform[41,42]. A possibility would be to use Dynamic Time Wrapping techniques, to match the tissue brain motion even under contraction/dilation of the ECG.

The use of DULM, and more generally ULM, is still limited by their need of relatively long acquisition time. The technique is inherently limited by its need for enough MBs flowing through every blood vessel. This process takes time, especially in small diameter vessels. Using motion-matching for time registration allows for a reduced waiting time compared to the previous need of an ECG-trigger, as depicted by the shaded grey areas in Fig 1. However, this improvement is minor compared to the time needed for the hardware to transfer the large amount of data needed for DULM and ULM. In the next decades, we could expect a large increase in transfer power, largely reducing the time needed to perform DULM and ULM.

## CONCLUSION

In conclusion, Dynamic Ultrasound Localization Microscopy (DULM) leads to better velocity estimation than the use of conventional Ultrasound Localization Microscopy (ULM). DULM can now be performed with a motion-matching time-registration on the tissue Doppler, allowing the use of DULM on any ULM dataset in a very computationally efficient manner. Pulsatility Index maps were extracted in a whole 2D-slice of a mouse brain. Dynamic flow parameters were successfully extracted on 2D and 3D dataset without the need of an ECG-trigger, as previously necessary for DULM. DULM could now be applied to every ULM application, even if only an average velocity map is needed.

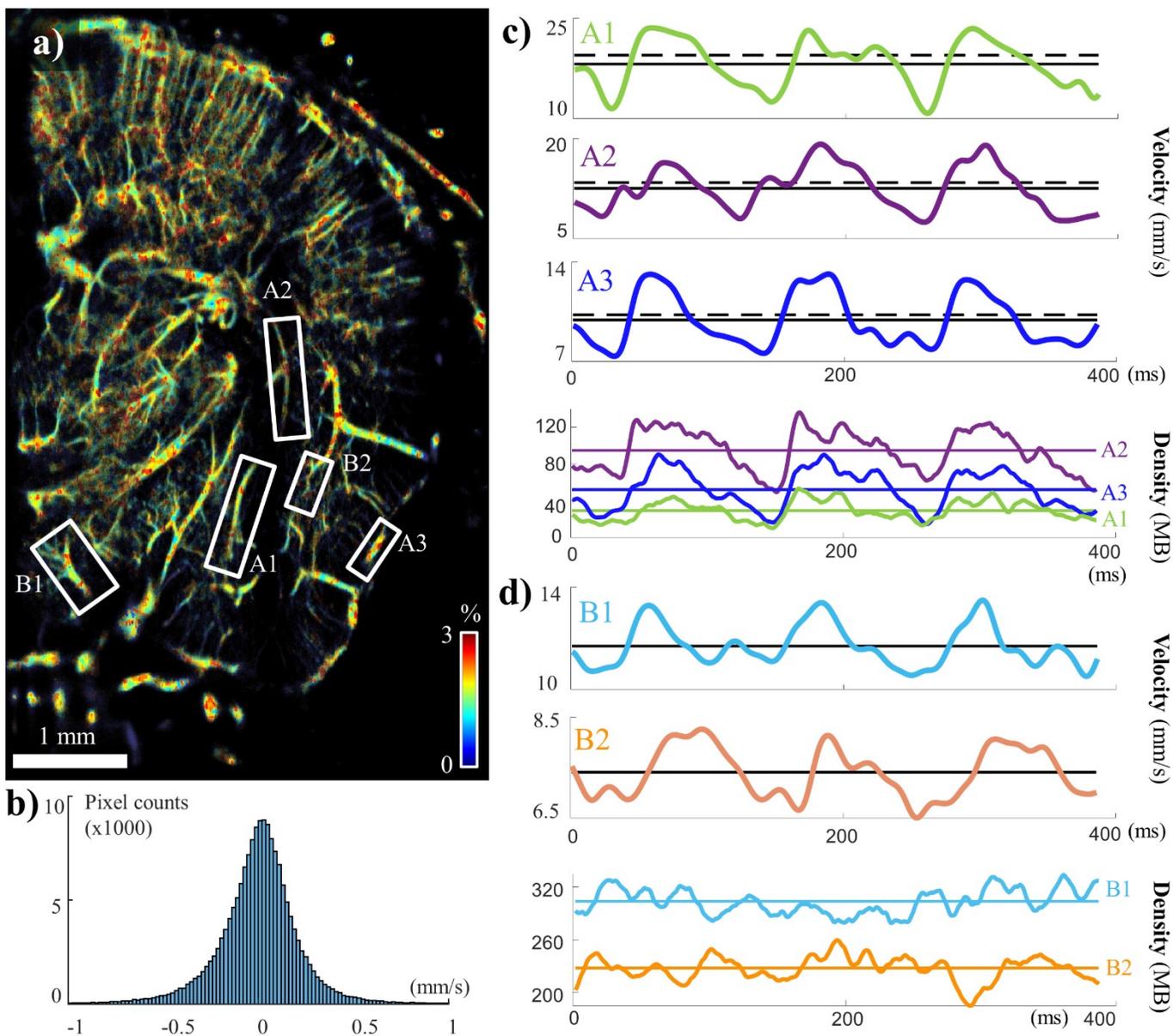

Figure 5: **Difference between the average velocity obtained through DULM and ULM process**. (a) represented in % of the DULM average velocity on a 10 µm×10 µm ~ λ/10×λ/10 grid. Only half of the brain is displayed here for easier representation (b) represented by a histogram in mm/s. (c-d) Change in velocity and microbubbles (MB) density along time in vessels of interests highlighted by white rectangles in (a). When the variation of MB density is significant compared to the mean MB density (c), the bias in mean velocity obtained with ULM (dotted black line) is also significant (5.9%, 4.5% and 2.8% of difference between DULM and ULM, for vessels numbered A1, A2 and A3 respectively). The mean velocity obtained with DULM is displayed as the continuous black line. When the variation of MB density is not significant compared to the mean MB density (c), the mean velocity obtained with DULM and ULM are similar with a difference < 0.2%.

FUNDINGS

This work received partial financial support from a variety of sources. It was supported by the Institute for Data Valorization (IVADO), the Canada Foundation for Innovation (Grant 38095), the Canadian Institutes of Health Research (CIHR, Grant 452530), and the New Frontiers in Research Fund (NFRFE-2018-01312). Further support came from the TransMedTech Institute, the Fonds de recherche du Québec — Nature et technologies, the Quebec Bio-Imaging Network, and the Canada First Research Excellence Fund (Apogée/CFREF). Additionally, computational resources were provided through the Digital Research Alliance of Canada.

DECLARATION OF GENERATIVE AI AND AI-ASSISTED TECHNOLOGIES IN THE WRITING PROCESS

During the preparation of this work the author(s) used OpenAI. (2023). ChatGPT (Sep 25 version) in order to improve readability and language. After using this tool/service, the author(s) reviewed and edited the content as needed and take(s) full responsibility for the content of the publication.

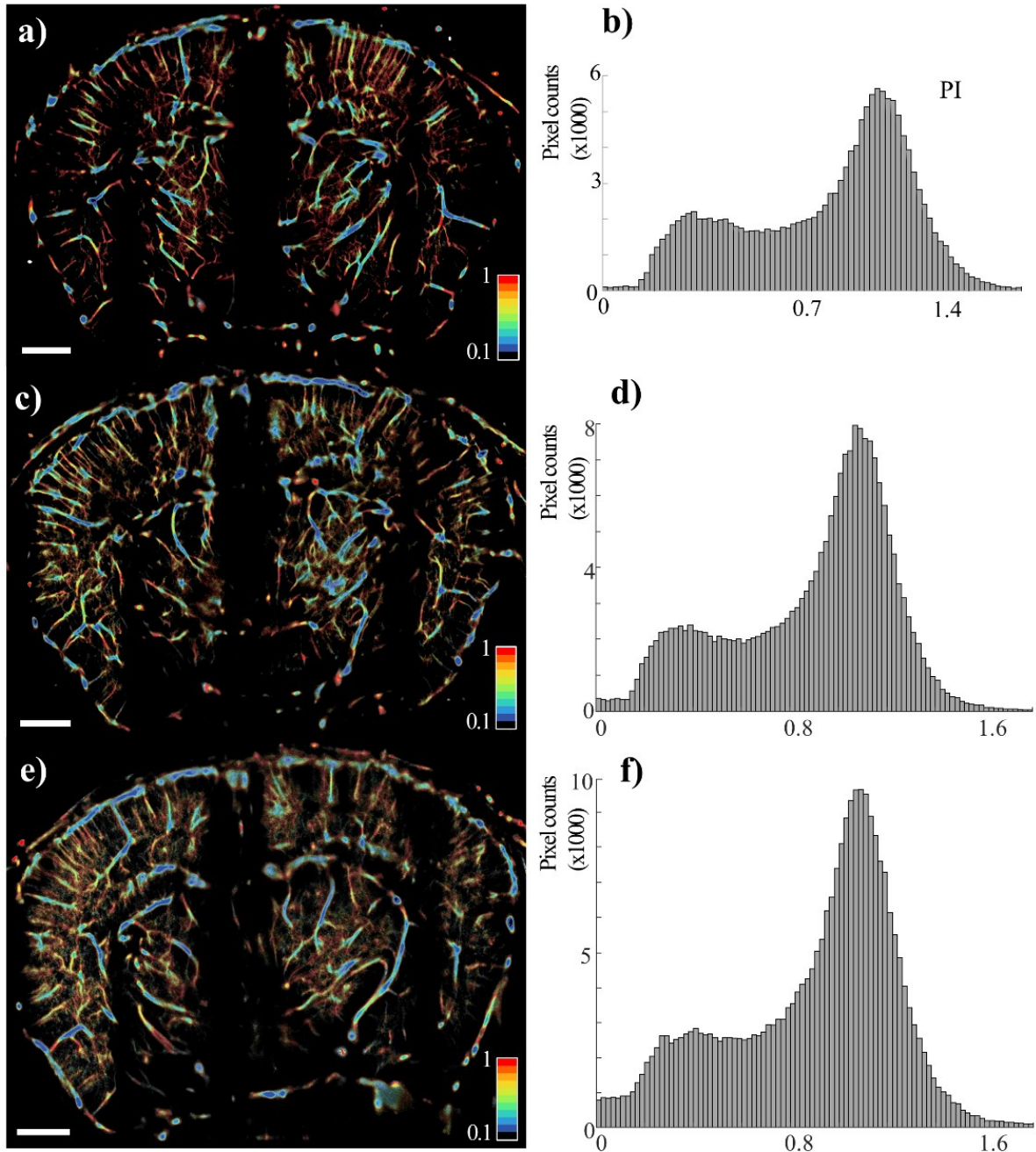

Figure 6: **Pulsatility Index (PI) Maps** (a-c-e) of the whole mice brain displayed on a 10 μm×10 μm ~ λ/10×λ/10 grid. The histograms of the resulting PI (b-d-f) show a bimodal distribution with both high and low PI pixels extracted.

REFERENCES

1. Gosling RG, King DH. The role of measurement in peripheral vascular surgery: arterial assessment by Doppler-shift ultrasound. *Proc R Soc Med*. 1974;67:447-449. doi:10.1177/00359157740676P113

2. Wagshul ME, Eide PK, Madsen JR. The pulsating brain: A review of experimental and clinical studies of intracranial pulsatility. *Fluids Barriers CNS*. 2011;8(1):1-23. doi:10.1186/2045-8118-8-5

3. Wielicka M, Neubauer-Geryk J, Kozera G, Bieniaszewski L. Clinical application of pulsatility index. *Medical Research Journal*. 2020;5(3):201-210. doi:10.5603/MRJ.a2020.0016

4. Aaslid R, Markwalder TM, Nornes H. Noninvasive transcranial Doppler ultrasound recording of flow velocity in basal cerebral arteries. *Journal of Neurosurgery*. 1982;57(6):769-774. doi:10.3171/jns.1982.57.6.0769

5. Lim EY, Yang DW, Cho AH, Shim YS. Cerebrovascular Hemodynamics on Transcranial Doppler Ultrasonography and Cognitive Decline in Mild Cognitive Impairment. Seo SW, ed. *JAD*. 2018;65(2):651-657. doi:10.3233/JAD-180026

6. de Roos A, van der Grond J, Mitchell G, Westenberg J. Magnetic Resonance Imaging of Cardiovascular Function and the Brain: Is Dementia a Cardiovascular-Driven Disease? *Circulation*. 2017;135(22):2178-2195. doi:10.1161/CIRCULATIONAHA.116.021978

7. Vikner T, Nyberg L, Holmgren M, Malm J, Eklund A, Wåhlin A. Characterizing pulsatility in distal cerebral arteries using 4D flow MRI. *J Cereb Blood Flow Metab*. 2020;40(12):2429-2440. doi:10.1177/0271678X19886667

8. Kim TN, Goodwill PW, Chen Y, et al. Line-Scanning Particle Image Velocimetry: An Optical Approach for Quantifying a Wide Range of Blood Flow Speeds in Live Animals. Secomb TW, ed. *PLoS ONE*. 2012;7(6):e38590. doi:10.1371/journal.pone.0038590

9. Meng G, Zhong J, Zhang Q, et al. Ultrafast two-photon fluorescence imaging of cerebral blood circulation in the mouse brain in vivo. *Proc Natl Acad Sci USA*. 2022;119(23):e2117346119. doi:10.1073/pnas.2117346119

10. Marchand PJ, Lu X, Zhang C, Lesage F. Validation of red blood cell flux and velocity estimations based on optical coherence tomography intensity fluctuations. *Sci Rep*. 2020;10(1):19584. doi:10.1038/s41598-020-76774-z

11. Walek KW, Stefan S, Lee JH, et al. Near-lifespan longitudinal tracking of brain microvascular morphology, topology, and flow in male mice. *Nat Commun*. 2023;14(1):2982. doi:10.1038/s41467-023-38609-z

12. Bercoff J, Montaldo G, Loupas T, et al. Ultrafast compound doppler imaging: providing full blood flow characterization. *IEEE Trans Ultrason, Ferroelect, Freq Contr*. 2011;58(1):134-147. doi:10.1109/TUFFC.2011.1780

13. Demené C, Pernot M, Biran V, et al. Ultrafast Doppler Reveals the Mapping of Cerebral Vascular Resistivity in Neonates. *J Cereb Blood Flow Metab*. 2014;34(6):1009-1017. doi:10.1038/jcbfm.2014.49

14. Errico C, Pierre J, Pezet S, et al. Ultrafast ultrasound localization microscopy for deep super-resolution vascular imaging. *Nature*. 2015;527(7579):499-502. doi:10.1038/nature16066

15. Couture O, Hingot V, Heiles B, Muleki-Seya P, Tanter M. Ultrasound Localization Microscopy and Super-Resolution: A State of the Art. *IEEE Trans Ultrason, Ferroelect, Freq Contr*. 2018;65(8):1304-1320. doi:10.1109/TUFFC.2018.2850811

16. Christensen-Jeffries K, Couture O, Dayton PA, et al. Super-resolution Ultrasound Imaging. *Ultrasound in Medicine & Biology*. 2020;46(4):865-891. doi:10.1016/j.ultrasmedbio.2019.11.013

17. Demené C, Robin J, Dizeux A, et al. Transcranial ultrafast ultrasound localization microscopy of brain vasculature in patients. *Nat Biomed Eng*. 2021;5(3):219-228. doi:10.1038/s41551-021-00697-x

18. Demeulenaere O, Bertolo A, Pezet S, et al. In vivo whole brain microvascular imaging in mice using transcranial 3D Ultrasound Localization Microscopy. *eBioMedicine*. 2022;79:103995. doi:10.1016/j.ebiom.2022.103995

19. Heiles B, Chavignon A, Hingot V, Lopez P, Teston E, Couture O. Performance benchmarking of microbubble-localization algorithms for ultrasound localization microscopy. *Nat Biomed Eng*. 2022;6(5):605-616. doi:10.1038/s41551-021-00824-8

20. Renaudin N, Demené C, Dizeux A, Ialy-Radio N, Pezet S, Tanter M. Functional ultrasound localization microscopy reveals brain-wide neurovascular activity on a microscopic scale. *Nat Methods*. 2022;19(8):1004-1012. doi:10.1038/s41592-022-01549-5

21. Heiles B, Chavignon A, Bergel A, et al. Volumetric Ultrasound Localization Microscopy of the Whole Rat Brain Microvasculature. *IEEE Open J Ultrason, Ferroelect, Freq Contr*. 2022;2:261-282. doi:10.1109/OJUFFC.2022.3214185

22. Bourquin C, Poree J, Lesage F, Provost J. In vivo pulsatility measurement of cerebral microcirculation in rodents using Dynamic Ultrasound Localization Microscopy. *IEEE Trans Med Imaging*. Published online 2021:1-12. doi:10.1109/TMI.2021.3123912

23. Bourquin C, Ikan L, Casanova C, Lesage F, Provost J. Quantitative pulsatility measurements using 3D Dynamic Ultrasound Localization Microscopy. *IEEE TRANSACTIONS ON MEDICAL IMAGING*. Published online 2023.

24. Cormier P, Poree J, Bourquin C, Provost J. Dynamic Myocardial Ultrasound Localization Angiography. *IEEE Trans Med Imaging*. 2021;40(12):3379-3388. doi:10.1109/TMI.2021.3086115

25. de Souza BOF, Cortes N, Casanova C. Pulvinar Modulates Contrast Responses in the Visual Cortex as a Function of Cortical Hierarchy. *Cerebral Cortex*. 2020;30(3):1068-1086. doi:10.1093/cercor/bhz149

26. Xing P, Porée J, Rauby B, et al. Phase Aberration Correction for in vivo Ultrasound Localization Microscopy Using a Spatiotemporal Complex-Valued Neural Network. Published online 2022. doi:10.48550/ARXIV.2209.10650

27. Demene C, Deffieux T, Pernot M, et al. Spatiotemporal Clutter Filtering of Ultrafast Ultrasound Data Highly Increases Doppler and fUltrasound Sensitivity. *IEEE Trans Med Imaging*. 2015;34(11):2271-2285. doi:10.1109/TMI.2015.2428634

28. Cigier A, Varray F, Garcia D. SIMUS: An open-source simulator for medical ultrasound imaging. Part II: Comparison with four simulators. *Computer Methods and Programs in Biomedicine*. 2022;220:106774. doi:10.1016/j.cmpb.2022.106774

29. Tinevez JY. A simple particle tracking algorithm for MATLAB that can deal with gaps. Published online 2019. https://github.com/tinevez/simpletracker

30. Provost J, Wei-Ning Lee, Fujikura K, Konofagou EE. Electromechanical Wave Imaging of Normal and Ischemic Hearts *In Vivo*. *IEEE Trans Med Imaging*. 2010;29(3):625-635. doi:10.1109/TMI.2009.2030186

31. Todd MM, Drummond JC. A Comparison of the Cerebrovascular and Metabolic Effects of Halothane and Isoflurane in the Cat. *Anesthesiology*. 1984;60(4):276-282. doi:10.1097/00000542-198404000-00002

32. Sullender CT, Richards LM, He F, Luan L, Dunn AK. Dynamics of isoflurane-induced vasodilation and blood flow of cerebral vasculature revealed by multi-exposure speckle imaging. *Journal of Neuroscience Methods*. 2022;366:109434. doi:10.1016/j.jneumeth.2021.109434

33. Song P, Trzasko JD, Manduca A, et al. Improved Super-Resolution Ultrasound Microvessel Imaging With Spatiotemporal Nonlocal Means Filtering and Bipartite Graph-Based Microbubble Tracking. *IEEE Trans Ultrason, Ferroelect, Freq Contr*. 2018;65(2):149-167. doi:10.1109/TUFFC.2017.2778941

34. Wiersma M, Heiles B, Kalisvaart D, Maresca D, Smith CS. Retrieving Pulsatility in Ultrasound Localization Microscopy. *IEEE Open J Ultrason, Ferroelect, Freq Contr*. 2022;2:283-298. doi:10.1109/OJUFFC.2022.3221354

35. Tang S, Song P, Trzasko JD, et al. Kalman Filter-Based Microbubble Tracking for Robust Super-Resolution Ultrasound Microvessel Imaging. *IEEE Trans Ultrason, Ferroelect, Freq Contr*. 2020;67(9):1738-1751. doi:10.1109/TUFFC.2020.2984384

36. Leconte A, Porée J, Rauby B, et al. A Tracking prior to Localization workflow for Ultrasound Localization Microscopy. Published online August 4, 2023. doi:10.48550/arXiv.2308.02724

37. Ohno Y, Adachi S, Motoyama A, et al. Multiphase ECG-triggered 3D contrast-enhanced MR angiography: utility for evaluation of hilar and mediastinal invasion of bronchogenic carcinoma. *Journal of Magnetic Resonance Imaging: An Official Journal of the International Society for Magnetic Resonance in Medicine*. 2001;13(2):215-224. doi:10.1002/1522-2586(200102)13:2<215::AID-JMRI1032>3.0.CO;2-2

38. Jian-Hung Liu, Geng-Shi Jeng, Tung-Ke Wu, Pai-Chi Li. ECG triggering and gating for ultrasonic small animal imaging. *IEEE Trans Ultrason, Ferroelect, Freq Contr*. 2006;53(9):1590-1596. doi:10.1109/TUFFC.2006.1678187

39. Fagman E, Perrotta S, Bech-Hanssen O, et al. ECG-gated computed tomography: a new role for patients with suspected aortic prosthetic valve endocarditis. *Eur Radiol*. 2012;22(11):2407-2414. doi:10.1007/s00330-012-2491-5

40. Richards LM, Towle EL, Fox DJ, Dunn AK. Intraoperative laser speckle contrast imaging with retrospective motion correction for quantitative assessment of cerebral blood flow. *Neurophoton*. 2014;1(01):1. doi:10.1117/1.NPh.1.1.015006

41. Mao S, Bakhsheshi H, Lu B, Liu SCK, Oudiz RJ, Budoff MJ. Effect of Electrocardiogram Triggering on Reproducibility of Coronary Artery Calcium Scoring. *Radiology*. 2001;220(3):707-711. doi:10.1148/radiol.2203001129

42. Kusumoto F. *ECG Interpretation: From Pathophysiology to Clinical Application*. 2nd ed. Springer Nature; 2020.